\documentclass{edp-conf}
\usepackage{graphicx}
\begin{document}

\TitreGlobal{SF2A 2003}

\title{Molecular gas and AGN fueling}
\author{Combes, F.}
\address{LERMA, Observatoire de Paris, 61 Av. de l'Observatoire, F-75014, Paris, France}
\runningtitle{Molecular gas and AGN fueling}
\setcounter{page}{237}
% Keep this line, even if the page will be settled afterwards..
\index{Combes, F.}

\maketitle
\begin{abstract} 
CO emission, tracing the molecular content and distribution 
in galaxies, is a privileged tool to trace gas
towards the nucleus, since the HI tracer is in general 
depleted there. A review  is done of recent CO line
observations, with sufficient spatial resolution to 
indicate the morphology and kinematics of the gas
near the nucleus. The puzzling result that nuclei 
presently observed in an active phase have little
sign of fueling, is discussed. 
\end{abstract}
%
%%-----------------------------
%%      your text
%%-----------------------------
\section{New high resolution maps}
  Recent surveys of CO molecules in nearby galaxies with millimeter
interferometers are coming up: the  BIMA SONG survey (Helfer et al. 2003, Sheth et al. 2002),
or the NUGA survey (Garc\'{i}a-Burillo et al. 2003, Combes et al. 2003). The
NUGA project consists in mapping at sub-arcsec resolution with the IRAM interferometer
more than a dozen of nearby active nuclei, with the goal to reveal the various gas morphologies
in the center, and determine the fueling mechanisms. One of the first striking results
of these surveys is the wide variety of dynamical perturbations in spiral galaxies, and
in particular asymmetries and lopsidedness, in addition to rings, spirals or bars, and 
embedded structures.

NGC 4303 (Sy2) has been mapped in CO at OVRO (Schinnerer et al 2002); it has a large
central mass concentration (large bulge), and follows the prototypes of double-bar
galaxies. The nuclear bar is encircled by the resonant ring of the primary bar. 
The bulk of the CO emission comes from two
straight gas lanes along the leading sides of the large-scale primary bar. Deviations
from circular velocities reach up to 90 km/s, along these gas lanes.
There is a conspicuous asymmetry in the center, where the CO gas emission is 
only weakly correlated to the UV continuum ring. 
The gas is not in the nuclear bar, nor clearly aligned with 
the ring (as in the similar types NGC 4314 or NGC 1068)
but follows the lanes. From the asymmetry in the CO map,
it is obvious that an m=1 self-gravitating mode is superposed to the double-bar
structure, and might play a role in the fueling.

NGC 5248 (late-type Sy2, Jogee et al. 2002) follows the prototype of a barred galaxy, with
low central mass concentration. The spiral structure extends from 10 kpc down to 100 pc.
There are at least two embedded structures, since inside the starburst ring 
of super-star clusters at 375 pc, develops another spiral structure with a second
H$\alpha$ ring inside, and a dust spiral revealed by HST.  The modelisation shows
that the gas is self-gravitating, even if the bar is driving the spiral density wave.
Although most of the central gas is confined to the nuclear
ring, there is some gas in the very center, maybe infalling intermittently.

The two galaxies NGC 3227 (Sy1.5) and NGC 1068 (Sy2), mapped in CO with IRAM interferometer,
reveal molecular nuclear disks in the center, with non-circular motions,
either due to a nuclear bar, or a warped disk.
The latter would explain the obscuration (molecular torus) required by the AGN type
(Schinnerer et al. 2000).
The two early-types Sy2 galaxies NGC 5728 and NGC 2273, mapped
with OVRO (Petitpas \& Wilson 2002) reveal very different gas morphologies, although they 
are both double-bar galaxies.
NGC 2273 shows evidence of a nuclear gas bar, aligned with the nuclear
stellar bar. Both are perpendicular to the primary bar, and therfore aligned on its x2 orbits.
On the contrary, NGC 5728 has no nuclear molecular bar but CO emission is tracing the 
spiral dust lanes, leading with respect to the primary bar (see also Combes \& Leon 2002).
This tends to suggest that the gas does not always follow the nuclear bar,
in all the phases of bar and double-bar episodes.

NGC 5005 (late-type LINER, Sakamoto et al 2000), mapped with OVRO, has a molecular gas distribution
concentrated in a nuclear disk and in an outside ring. Non-circular motions show that the
gas is following x1 orbits in the nuclear bar. A light stream of gas reveals epidodic accretion from
the ring towards the center.  This transient (intermittent) gas flow towards the center
could be the way fueling is regulated towards central starburst and active nuckei.
With even more spatial resolution, dust mapped with HST in the central kpc of galaxies, reveal 
statistically a large number of nuclear spirals, and sometimes nuclear bars
(e.g. Pogge \& Martini 2002). Decoupled
nuclear disks are therefore quite frequent (dust, gas, young stars)
as the result of bars and interactions.

\section{Various possibilities for the fueling models}

Models of gas flow in barred (and double-barred)
galaxies are compatible with observations.
Depending on the sound speed (more or less velocity dispersion in the gas)
gas waves develop with more or less pitch angle inside ILR.
In the very nucleus, inside the resonant ring of the primary bar,
there is a region of 
low bar torque (x2 orbits are nearly circular), and no shocks.
The fueling might be due to other perturbations than the 
traditional m=2 ones.
It is possible that an m=1 asymmetry, superposed to the m=2 perturbations,
drives the gas towards the center. This could be triggered by
asymmetric gas accretion.

Molecular gas observations tend to show that the
fueling mechanism is not only the bar-within-bar one, 
but molecular clouds are able to infall, even in the absence 
of strong torque, transiently inside the nuclear ring. 
It could be through asymmetries, or also through dynamical 
friction.

\section{A special axisymmetric case: NGC 7217}

NGC 7217 is an early-type Seyfert/Liner galaxy,
isolated, and with a remarkably regular morphology. It
possess 3 rings at 10.7, 32 and 77 ``  (0.75, 2.2 and 5.4 kpc),
interpreted as nuclear ring, inner and outer rings 
of an oval distortion (Buta et al. 1995).
The blue color of the rings indicates star formation.
NGC 7217 has a conspicuous bulge dominating the luminosity
of the galaxy.
The center of the galaxy has been mapped with the IRAM interferometer
(cf Fig. 1). The molecular gas is concentrated in the nuclear
ring, which has an abrupt profile towards the center,
delineating a real hole in the gas. The molecular gas 
then follows the flocculent spiral arms that
begin just outside the ring. 
The dust mini-spiral inside the ring (Fig. 1) has a counterpart in
CO emission, much better seen in the CO(2-1) line.
This gas lane corresponds to a spiral winding down from
the northern part of the ring, as if some gas was infalling from
the ring towards the center. 

\begin{figure}
\centering
\rotatebox{-90}{\includegraphics[width=6cm]{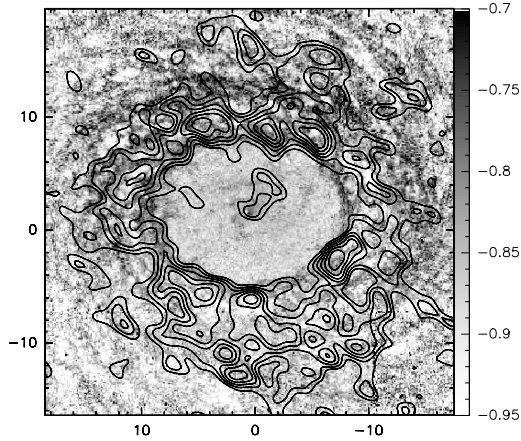}}
\caption{ NGC 7217 CO(1-0) contours superposed on the HST $V - I$ color image.
 Darker pixels correspond to redder colors (Combes et al. 2003).}
\label{fig1}
\end{figure}

N-body simulations have been run, to understand the
role of the dynamical components 
(bulge, dark matter and disk)
and that of physical processes (gas self-gravity, star-formation
and feedback) in the gas morphology and possible
fueling. A polar grid particle-mesh code with variable resolution
(from 17pc to 1kpc) has been used,
taking into account self-gravity of the gaseous and stellar
disk, gas dissipation via a sticky particle
algorithm, star-formation and feedback.  The observed rotation
curve is compatible with two extreme models: a maximum
disk, or maximum bulge model. In the maximum disk hypothesis,
a strong bar develops, and must have disappeared
today; but the bar destruction requires
at least 5\% of the total mass in gas, which is more than 10 times
the gas present today. The maximum bulge model
yields the best fit to the data. Only an oval distortion develops,
since the disk is stabilised by the bulge. The oval distortion
is sufficient to generate the resonant rings in the disk, if the initial
gas amount is about twice than today. The 
star formation consumes this gas away. 
The oval distortion, still present today, maintains by its torques the gas hole in the
center and the abrupt density profile inside the ring (Fig. 2).

\begin{figure}
\centering
\rotatebox{-90}{\includegraphics[width=3cm]{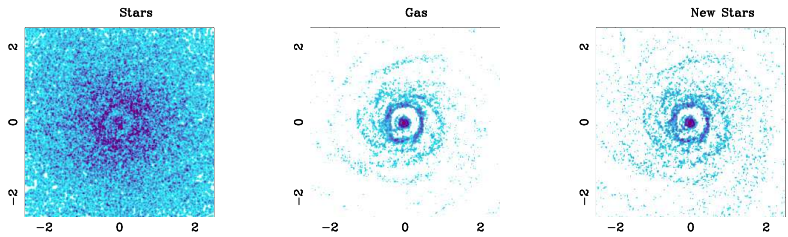}}
\caption{ {\bf Left:} Stellar component
distribution, in the bulge-dominated model of NGC 7217
{\bf Middle:}  Gaseous component
{\bf Right:} Young stars formed during the simulation}
\label{fig2}
\end{figure}

\section{Conclusions}

Molecular gas in the central parts of galaxies reveal various 
morphologies:  nuclear rings, disks, nuclear spiral arms, nuclear bars.
The fueling mechanisms are varied also: m=2 density waves and torques
but also viscosity, dynamical friction, and m=1 asymmetries.
Maps show much more perturbations and asymmetries than found
in models. Self-gravity of the gas appears important.
The dynamical  mechanisms are different according to the 
bulge-to-disk mass ratio: late-type galaxies have only
one bar, and spirals winding all the way through, while
early-type galaxies possess decoupled nuclear bars, with
highly different time-scales between nuclear and large-scale features.

%%-----------------------------
%%      your bibliography
%%-----------------------------

\end{document}